# The influence of out-of-plane disorder on the formation of pseudogap and Fermi arc in $Bi_2Sr_{2-x}R_xCuO_y$ (*R*=La and Eu)

Y. Okada[1], T. Takeuchi[2], A. Shimoyamada[3], S. Shin[3], H. Ikuta[1]

[1]*Department of Crystalline Materials Science, Nagoya University, Nagoya 464-8603, Japan*

[2]*EcoTopia Science Institute, Nagoya University, Nagoya 464-8603, Japan*

[3]*Institute for Solid State Physics (ISSP), University of Tokyo, Kashiwa 277-8581, Japan*

**Abstract**

We found that the length of the Fermi arc decreases with increasing out-of-plane disorder by performing angle resolved photoemission spectroscopy (ARPES) measurements in the superconducting state of optimally doped *R*=La and Eu samples of $Bi_2Sr_{2-x}R_xCuO_y$. Since out-of-plane disorder stabilizes the antinodal pseudogap as was shown in our previous study of the normal state, the present results indicate that this antinodal pseudogap persists into the superconducting state and decreases the Fermi arc length. We think that the shrinkage of the Fermi arc reduces the superfluid density, which explains the large suppression of the superconducting transition temperature when out-of-plane disorder is increased.

**Introduction**

Recently, Fujita *et al.* reported that out-of-plane disorder introduced through a replacement of Sr by a rare earth element $R$ in the $Bi_2Sr_{2-x}R_xCuO_y$ system suppressed significantly the transition temperature $T_c$, although the increase of the residual in-plane resistivity was relatively small [1]. They stressed the importance and novelty of the effect of out-of-plane disorder as their results imply that this type of disorder influences $T_c$ without being a strong scatterer. The understanding of the influence of out-of-plane disorder is important since all high-$T_c$ cuprates are suffered more or less from disorder that is inherently introduced with carrier doping. Extensive studies along this line have been mostly conducted for the $La_2CuO_4$ family [2-5]. For this material, $T_c$ depends on the size of the cation that substitutes for La, and the connection between $T_c$ suppression and the structural change have been well studied. Information about the influence on the electronic structure by out-of-plane disorder is, however, rather limited, probably because of the difficulty to apply surface sensitive experimental techniques to the $La_2CuO_4$ family. In contrast, $Bi_2Sr_{2-x}R_xCuO_y$ has a great advantage that a clean surface can be easily prepared by cleaving. In this system, the degree of out-of-plane disorder is controlled through the mismatch of the ionic radius between Sr and the $R$ element [6,7]. It is not easy to connect the structural change and disorder because of the nonstoichiometry and superstructure [8], but the electronic structure in both momentum and real spaces can be studied with the same material via angle resolved photoemission spectroscopy (ARPES) [9,10] and scanning tunneling microscopy/spectroscopy (STM/STS) techniques [11-13].

Some of the present authors have extensively studied the $Bi_2Sr_{2-x}R_xCuO_y$ system using single crystals with varying $x$ over a wide range for $R$=La, Sm, and Eu [14]. The results clearly show that $T_c$ at the optimal doping $T_c^{max}$ depends strongly on the $R$ element and decreases with increasing out-of-plane disorder. They also found that the carrier range where superconductivity takes place on the phase diagram becomes narrower with increasing out-of-plane disorder [14]. To elucidate the origin of this strong $R$ dependence, we have studied the electronic structure by means of ARPES measurements [15,16]. We focused in the previous studies on the so-called antinodal position of the Fermi surface and on the normal state, and found that the temperature $T^*$, above which the pseudogap structure at the Fermi energy $E_F$ disappears, increases with disorder when the doping is the same. Because $T_c$ decreases with increasing disorder [14], the suppression of $T_c$ correlates well with the stabilization of the pseudogap at the antinodal region, implying that the antinodal pseudogap state competes with high-$T_c$ superconductivity. The question to be arisen then is why the antinodal pseudogap suppresses high-$T_c$ superconductivity. It was not possible to experimentally address to this problem because the measurements in our previous works were restricted to the antinodal region. Therefore, we extended our study and measured the electronic structure along the whole Fermi surface of optimally doped $R$=La and Eu samples of $Bi_2Sr_{2-x}R_xCuO_y$, i.e., samples that have similar doping but a different degree of out-of-plane disorder.

**Experimental**

The optimally doped single crystals of $Bi_2Sr_{2-x}R_xCuO_y$ with $R$=La and Eu were grown from polycrystalline rods that had a nominal composition of $Bi_2Sr_{1.65}R_{0.35}CuO_y$ ($R$=La and Eu) with the same growth condition reported previously [14]. $T_c$ of the samples were 33 K for $R$=La and 18 K for $R$=Eu, respectively, which correspond to their $T_c^{max}$ values, and indicate that they are optimally doped [14]. High resolution ARPES spectra were accumulated using a Scienta SES2002 hemispherical analyzer with the Gammadata VUV5010 photon source (He I$\alpha$) at the Institute of Solid State Physics (ISSP). The energy resolution was about 10 meV and the samples were cleaved in a pressure better than $7 \times 10^{-11}$ Torr. All the measurements in this report were performed at 8 K, which is low enough compared to $T_c$ for both samples.

**Results and Discussion**

Figs. 1(a) and (b) are intensity plots of the ARPES spectra integrated within ±10 meV around $E_F$ for the $R$=La and Eu samples, respectively. The Fermi momentum $k_F$ deduced from the peak position of the momentum distribution curves (MDCs) is plotted in fig. 1(c). First of all, we note from fig. 1(c) that the size of the Fermi surface of the two samples is identical within the experimental error, indicating that the carrier concentration of these two samples is very similar. Fig. 1(d) shows the temperature dependence of thermopower around room temperature. The thermopower at room temperature has a strong correlation with hole doping [17,18], and we see that the two samples exhibited very similar thermopower, also suggesting that the carrier concentration of the two samples is very similar. Hence, both the size of Fermi surface and the room temperature value of thermopower consistently indicate that the doping level of the two samples is similar, and the electronic structure discussed below stems mostly from the change in the degree of out-of-plane disorder at a fixed hole doping.

As mentioned already, the $T_c$ values of the present samples correspond to their $T_c^{max}$ values, indicating that they are optimally doped. Nevertheless, one may notice that the Fermi surface of the two samples of fig. 1(c) is rather large. The solid line in fig. 1(c) corresponds to the Fermi surface calculated by assuming a hole doping of $p$=0.21 and that the Luttinger sum rule holds. The assumed hole doping is certainly larger than the optimal doping of many other high-$T_c$ cuprates [17-19]. It is, however, consistent with the report that the superconducting region of the $Bi_2Sr_2CuO_y$ (Bi2201) family locates at the more overdoped side on the phase diagram when compared to other high-$T_c$ cuprates [20]. Further, the thermopower at room temperature shown in fig. 1(d) is smaller than most of other optimally doped high-$T_c$ superconductors [21], which implies that the hole concentration is larger, consistent with the large Fermi surface. To discuss the difference of the carrier range where superconductivity occurs compared with other high-$T_c$ cuprates is beyond the scope of this report, but may be related with the relatively low $T_c$ of Bi2201. We may stress, however, that the close correspondence between the size of Fermi surface and thermopower indicates that the thermopower value at room temperature is a good measure of doping, although the connection between these two quantities is merely empirical and has not been justified from a basic physics viewpoint [22,23].

As shown in figs. 1(a) and (b), the mapping of integrated intensity of the ARPES spectra resulted in a disconnected arc-like feature, similar to the one that had been seen in other high-$T_c$ cuprates [24-28]. The length of this so-called Fermi arc is obviously shorter for the $R$=Eu sample (fig. 1(b)). In figs. 2(a) and (b), we show the ARPES spectra of the two samples along the Fermi surface for several momentum points. Clear quasiparticle peaks were observed in the spectra taken near the nodal point for both samples. On the contrary, the spectra at the antinodal region exhibited no clear peaks. Such dichotomy of the ARPES spectra has been widely observed in high-$T_c$ cuprates [29] and is closely related to the formation of a Fermi arc [24-28]. We plotted in fig. 3 the momentum where a quasiparticle or a coherence peak was observed in the ARPES spectrum by solid symbols. It can be seen that the momentum region where the peaks were detected is narrower for the $R$=Eu sample, i.e. when the out-of-plane disorder is larger. Hence we can conclude that the Fermi arc, which we define as the region where one can observe a clear peak in the ARPES spectrum in the superconducting state, shrinks with increasing out-of-plane disorder.

As we have reported previously, the pseudogap that is formed at the antinodal region is stabilized with increasing out-of-plane disorder [14,15]. In the present work, on the other hand, we observed that with the increase of out-of-plane disorder, the length of the Fermi arc decreased, or in other words, a larger portion of the Fermi surface vanished. The ARPES measurements of the present work were performed at 8 K, i.e., the spectra were taken at the superconducting state. It is quite natural therefore to conclude that the pseudogap formed at the antinodal region is robust and persists down to the superconducting state, and is the reason of the disconnected arc feature of the Fermi surface. If we then assume that only the states on the Fermi arc can participate to superconductivity, the superfluid density would decrease with the increase of out-of-plane disorder. The suppression of $T_c$ observed with the increase of out-of-plane disorder can hence be easily understood based on the Uemura relation [30], which suggests that $T_c$ is a function of superfluid density.

In a recent paper, Tanaka *et al.* reported that the length of the Fermi arc of $Bi_2Sr_2CaCu_2O_y$ (Bi2212) determined by the same procedure as fig. 3 shrinks with decreasing hole doping [28]. They have also shown that the energy gap at the antinodal region, presumably the pseudogap, increases with hole underdoping. Therefore, the shrinkage of the Fermi arc with the stabilization of the antinodal pseudogap is a widely observed for high-$T_c$ cuprates [24-28,31-33]. Our results add a new piece of information and indicate that the Fermi arc shrinks even at the same hole doping when the antinodal pseudogap is stabilized by out-of-plane disorder.

In a recent STM/STS work, Sugimoto *et al.* reported an increase in the spatially averaged energy gap with increasing out-of-plane disorder [34]. They also observed a decrease in the volume fraction of the superconducting region that can be characterized by a small energy gap and coherence peaks of the STS spectra with increasing out-of-plane disorder. Our ARPES data complement the STM/STS observations and show that the decrease of the superconducting volume fraction in the real space corresponds to a shrinking of the Fermi arc in the momentum space. This may give a clue to connect between the ARPES data and the inhomogeneity of the electronic structure seen in the STM/STS experiments. It is worthwhile noting here that a very recent STM/STS experiment unveiled a new narrow homogeneous gap that is superimposed on the inhomogeneous and broad gap seen before [35-37]. The homogeneous gap, which was assigned to the superconducting gap because it disappears near $T_c$ [37], was argued to correspond to the energy gap that opens near the node, or in other words, on the Fermi arc.

As argued so far, we think that the most relevant parameter that determines $T_c$ when out-of-plane disorder is changed at a fixed hole doping is the amount of carrier that can participate to superconductivity as reflected in the length of the Fermi arc. However, other parameters that may also affect $T_c$ require consideration to draw a firm conclusion. In particular, the size of the superconducting gap that is formed on the Fermi arc may be an important parameter [38]. In the present work, it was not possible to resolve the superconducting gap clearly in our nodal spectra because $T_c$, and consequently the size of the superconducting gap, is small in Bi2201. Laser based ARPES, which has a much better energy and momentum resolution [39,40], would be a powerful tool to address the influence of out-of-plane disorder on the superconducting gap in subsequent studies.

**Conclusion**

We studied the electronic structure of optimally doped $Bi_2Sr_{2-x}R_xCuO_y$ with $R$=La and Eu by means of ARPES measurements. We found that the length of the Fermi arc in the superconducting state shrinks with increasing out-of-plane disorder when doping is the same. This suggests that the $T_c$ suppression with increasing out-of-plane disorder is due to the reduction of the states on the Fermi surface participating to superconductivity, or in other words, to the reduction of the superfluid density.

**Acknowledgement**

We would like to thank S. Kuno and H. Komoto of Nagoya University for experimental assistance.

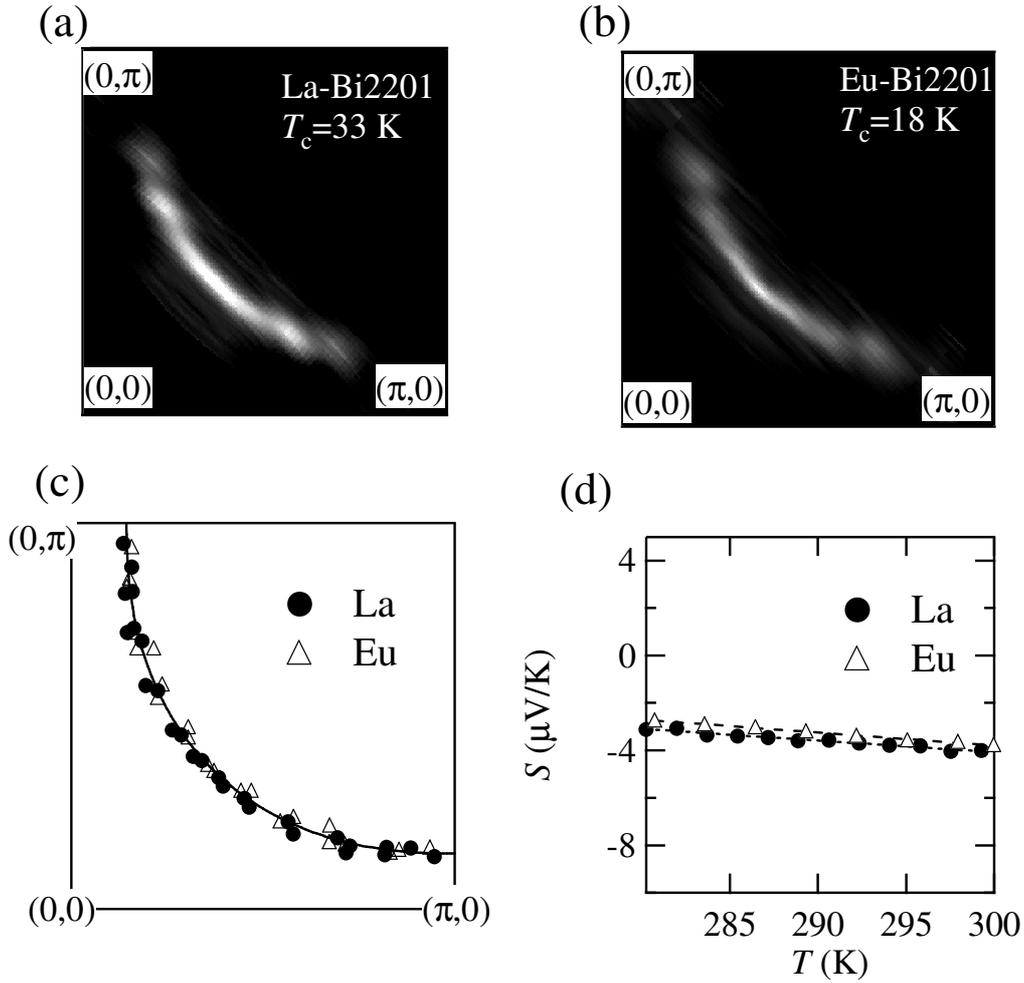

Fig. 1 Intensity plot of ARPES spectra integrated within ±10 meV around the Fermi energy for optimally doped $Bi_2Sr_{2-x}R_xCuO_y$ with $R$=La (a) and $R$=Eu (b). The intensity due to the subband is cut off by the scale. (c) Plot of the Fermi momentum $k_F$ determined from momentum distribution curves (MDCs). The solid line corresponds to the Fermi surface calculated by assuming a hole doping of $p$=0.21 and that the Luttinger sum rule holds. (d) The temperature dependence of thermopower of the two samples around room temperature.

Y. Okada *et al.*

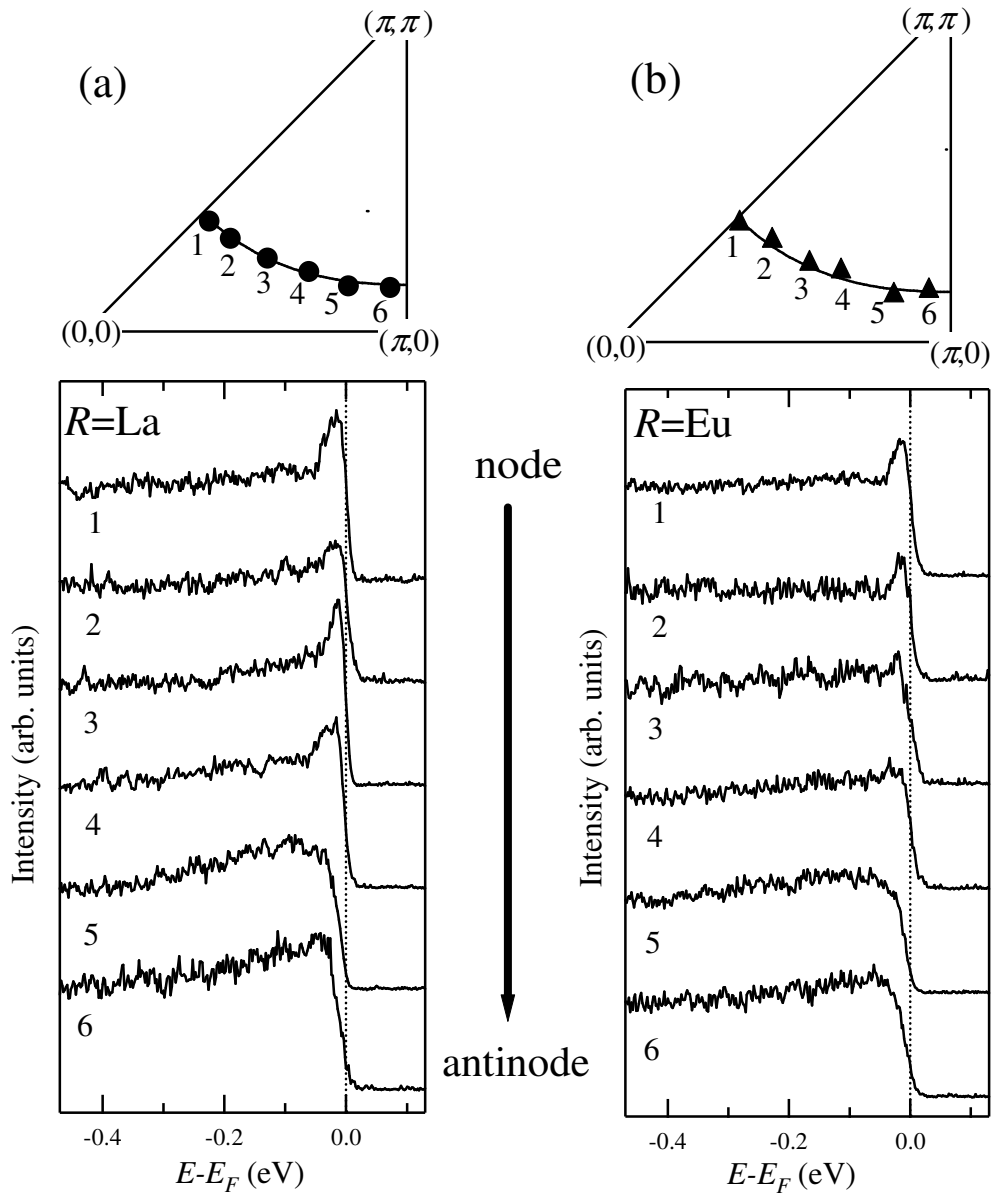

Fig. 2 Spectra taken at various momentum points along the Fermi surface of the optimally doped (a) $R$=La and (b) $R$=Eu samples of $Bi_2Sr_{2-x}R_xCuO_y$ measured at 8 K. All the spectra shown here were taken in the first Brillouin zone. The momentum points where each spectrum was taken are sketched in the upper part of each figure.

Y. Okada *et al.*

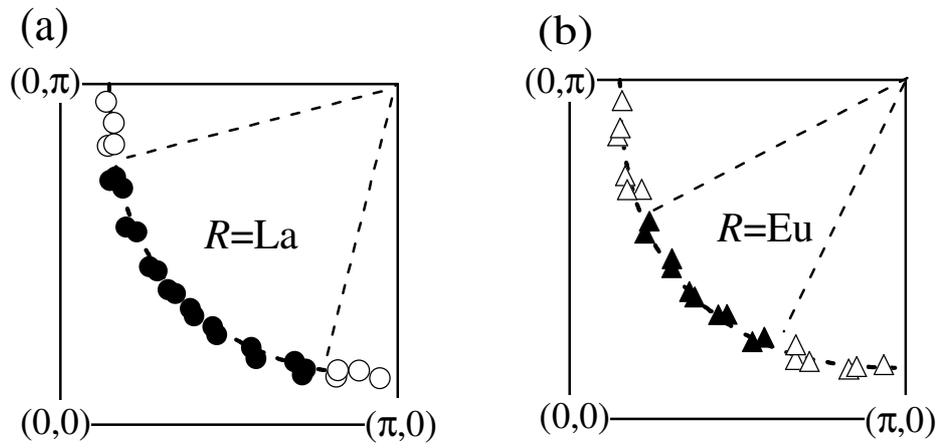

Fig. 3 The momentum region where quasiparticle or coherence peaks were resolved for (a) the $R$=La and (b) the $R$=Eu sample. Filled markers represent the momentum points where a quasiparticle or a coherence peak was observed while open markers where no peak was observed. Note here that these data were obtained at the superconducting state.

Y. Okada *et al.*